\documentstyle[12pt]{article}
\newcommand{\be}{\begin{equation}}
\newcommand{\ee}{\end{equation}}
\newcommand{\bea}{\begin{eqnarray}}
\newcommand{\eea}{\end{eqnarray}}
\newcommand{\bi}{\begin{itemize}}
\newcommand{\ei}{\end{itemize}}
\newcommand{\bc}{\begin{center}}
\newcommand{\ec}{\end{center}}
\newcommand{\bfl}{\begin{flushleft}}
\newcommand{\efl}{\end{flushleft}}
\newcommand{\bfr}{\begin{flushright}}
\newcommand{\efr}{\end{flushright}}

\def\6{\partial} \def\a{\alpha} \def\b{\beta}
\def\g{\gamma} \def\d{\delta} 
\def\e{\epsilon}
  \def\th{\theta}
  \def\l{\lambda}
\def\m{\mu} \def\n{\nu} \def\x{\xi} 
\def\r{\rho} \def\s{\sigma} 
  
\def\o{\omega} \def\G{\Gamma} 
  \def\S{\Sigma}
  \def\O{\Omega}

\def\={\!\!\!&=&\!\!\!}
\def\+{\!\!\!&&\!\!\!+~}
\def\-{\!\!\!&&\!\!\!-~}

\newcommand{\DD}{{\cal D}}

\newcommand{\FF}{{\cal F}}

\newcommand{\TT}{{\cal T}}

\newcommand{\journal}[4]{{\em #1~}#2\,(19#3)\,#4;}

\newcommand{\pr}{\journal {Phys. Rev.}}

\newcommand{\prl}{\journal {Phys. Rev. Lett.}}
\newcommand{\jmp}{\journal {J. Math. Phys.}}
\newcommand{\rmp}{\journal {Rev. Mod. Phys.}}
\newcommand{\cmp}{\journal {Comm. Math. Phys.}}

\newcommand{\np}{\journal {Nucl. Phys.}}

\begin{document}
\title{Observables of Euclidean Supergravity}
\author{
{\em Ion V. Vancea,}\\
{\em Department of Theoretical Physics, Babes-Bolyai University of Cluj,}\\
{\em Str. M. Kogalniceanu Nr.1, RO-3400 Cluj-Napoca, Romania}}

\date{\today}
\maketitle

\begin{abstract}
The set of constraints under which the eigenvalues of the Dirac 
operator can play the role of the dynamical variables for Euclidean 
supergravity is derived. These constraints arise when the gauge 
invariance of the eigenvalues of the Dirac operator is imposed. They 
impose conditions which restrict the eigenspinors 
of the Dirac operator. 
\end{abstract}
PACS 04.60.-m, 04.65.+e
\thispagestyle{empty}
\newpage

During the last  years, the Dirac operator has become a very powerful
tool for studying the geometrical properties of the manifolds as
well as the fundamental physics that takes place on them. Not long
ago, Connes showed, in the context of noncommutative geometry, that 
the Dirac operator contains full information about the geometry of 
space-time \cite{co}. It turned out that this property makes the Dirac
operator suitable for describing the dynamics of the general 
relativity and that, at least in principle, the Dirac operator can be 
used instead of the metric to describe the gravitational field 
\cite{ct}-\cite{cff}. However, there are still some major 
problems that must be solved before this point of view be
totally accepted. One of the most important ones is raised by
the fact that the spectrum of the Laplace type operators
(like the squared Dirac operator) cannot uniquely determine the topology
and the geometry of a four-dimensional Riemannian manifold (a detailed analysis
on this topic can be found in \cite{pbg}). Another problem arises
when Riemannian manifolds without boundary are considered. They provide
only an idealization of the manifolds encountered in physics since
boundary terms play a crucial role in many important physical
phenomena, as for example in determining  the black-hole
entropy from a perturbative evaluation of the path-integral
for the partition function \cite{gghh}. 
In the spectral geometry approach to Euclidean gravity, 
a major role is played by the eigenvalues $\l^{n}$'s of the Dirac 
operator $D$ which are 
diffeomorphism-invariant functions of the geometry and thus can be 
considered as the observables of general relativity. In a very 
recent paper, Landi and Rovelli expressed the Poisson bracket of
$\l^n$'s in terms of the components of the energy-momentum tensor
of the corresponding eigenspinor, and derived the Einstein equations
from a spectral action with no cosmological term \cite{lr}. This 
could be a new way to think of quantum gravity, but this method 
works at the moment being only for the
Euclidean theory. There are, however, several attempts to implement 
the method in the Lorentzian case \cite{mk,eh}. Nevertheless, the
Euclidean case by itself is quite interesting and it is the only one  
for which the ellipticity of the some differential operators can 
be ensured. Also, many interesting problems can be formulated in a well 
defined form only in the Euclidean setting. In gravity, the  
Euclidean case allows to derive 
several interesting solutions of the Einstein equations, as Euclidean 
wormholes which, together with the minisuperspace technique, were
invoked to explain constants in nature, as the vanishing cosmological
constant \cite{sc}. Another interesting problem is that in some 
circumstances spacetime might change signature and become Euclidean.
That shows that it would be worthwhile to investigate what happens 
when physical objects pass from one region to the other one. These
examples motivate enough the study of the Euclidean systems and in
this context it is natural to ask whether the description of the 
space-time by means of the eigenvalues of the Dirac operator can be 
extended further to include the supersymmetric case. If a supersymmetric
partner of the metric is considered, and the local supersymmetry is imposed,
we are led to Euclidean supergravity. This kind of systems has been
extensively studied lately mainly into the frame of path integral 
quantization of supergravity with a stress on the problem of the
boundary conditions which are to be imposed on the fermions \cite{de,eeo}
(see also \cite{bge}). In this paper it is addressed the question whether 
the eigenvalues of the Dirac operator can be used as observables for
Euclidean supergravity.

Consider Euclidean minimal supergravity on a compact D=4 (spin) manifold
with no boundary. The graviton is represented in the tetrad formalism by
the fields $e^{a}_{\m}(x)$; $\m =1,\cdots,4 $ are space-time indices and
$a=1,\cdots,4$ are internal Euclidean indices, raised and lowered by the
Euclidean metric $\d _{ab}$. The metric field is 
$g_{\m \n}(x) = e_{\m}^{a}(x) e_{\n a}(x)$. The gravitino is represented
by a Euclidean spin-vector field $\psi_{\m}(x)$. In order to have 
" Euclidean Majorana spinors" and to maintain the correct number of
degrees of freedom required by a supersymmetric theory, the adjoint spinors
are defined via Majorana conjugation relation $\bar{\psi} = \psi^{T}C$.
This encounters the problem posed by the fact that there is no Majorana
spinor representation of $SO(4)$ and ensures the theory is supersymmetric  
\cite{pn}( for other discussions on the "Euclidean Majorana spinors"
see also \cite{ge,kt,bge}). The spin connection $\o_{\m ab}(e,\psi )$ is 
defined as
\be
\o_{\m ab}(e,\psi ) =  \stackrel{\circ}{\o_{\m ab}}(e) + K_{\m ab}(\psi )   
\label{conn}
\ee
where $  \stackrel{\circ}{\o_{\m ab}}(e) $ is the usual spin-connection of 
gravity. In units such that $8\pi G=1 $ the following relations are assumed 
true
\bea
\stackrel{\circ }{\o_{\m ab} } (e) & = &  \frac{1}{2} e_{a}^{\m}(\6_{\m}
e_{b\n}-
\6_{\n}e_{b\m}) +\frac{1}{2}e_{a}^{\r}e_{b}^{\s}\6_{\s}e_{\r c}e^{c}_{\m} -
(a\leftrightarrow b) \\
K_{\m ab}(\psi ) &=& \frac{i}{4}(\bar{\psi_{\m}}\g_{a}\psi_{b} - 
\bar{\psi_{\m}}\g_{b}\psi_{a} + \bar{\psi_{b}}\g_{\m}\psi_{a}). \label{kap}
\eea
As usual in the supersymmetric case, there are two covariant derivative 
acting on space-time tensors and space-time spinors, respectively. The 
minimal covariant derivative, when acts on vectors, for example, 
is expressed in terms of Christoffel symbols as
\be
D_{\m}V^{\n} = \6_{\m}V^{\n} + \G^{\n}_{\m \s}V^{\s}
\ee
and the non-minimal covariant derivative which acts on spinors
reads as follows
\be
\DD_{\m} \phi = \6_{\m} \phi -\frac{i}{2}\o_{\m ab}(e,\psi ) \s^{ab}\phi 
\label{nmd}
\ee
where $\s^{ab} = \frac{1}{4} [\g^{a},\g^{b}] = i\S^{ab}$ and $\g^{a}$'s
form an Euclidean representation of the Clifford algebra $C_4$:
$\{ \g^{a},\g^{b} \} = \d^{ab}$.

The phase space of the theory is defined as the space of the solutions of the
equations of motion, modulo the gauge transformations \cite{lr,cm}. The gauge
transformations are 4D diffeomorphisms, local $SO(4)$ rotations and the local
$N=1$ supersymmetry. Then the phase space, which is covariantly defined, is
the space of all $e, \psi$ that are solutions of the equations of motion
modulo diff's, internal rotations and local supersymmetry. Because the phase
space is defined over the solutions of the equations of motion it is sufficient
to consider only on-shell supersymmetry. In this case, the supersymmetric
algebra closes over only graviton and gravitino. Off-shell, the supersymmetric
algebra usually requires six more bosonic fields since there is a mismatch of
the fermionic and bosonic degrees of freedom. As in \cite{lr} the observables 
of the theory are the functions on the phase space. For Euclidean gravity
the eigenvalues $\l^n$'s of $D$ define a discrete real family of real-valued
functions on the space of all tetrads and for every $n$ the function 
$\l^{n}(e)$ is invariant under diff's and under internal rotations. Therefore
they are well defined on the phase space and they are observables of general
relativity. For the present case one has to find what are the conditions
under which $\l^n$'s are gauge invariant such that they can be used as 
observables of supergravity. The presence of the gravitino and the 
requirement of the local susy give a non-trivial solution to this problem.

In the present case the Dirac operator is defined as
\be
D = i\g^a e_{a}^{\m} (\6_{\m} + \o_{\m bc}(e,\psi ) \g^b \g^c ) \label{dirop}
\ee
and it acts on the Euclidean spinors defined on the manifold. It is possible 
to construct the Dirac operator such that it is self-adjoint on the Hilbert space 
of spinor fields with the scalar product defined as in \cite{lr}
\be
<\psi ,\phi > = \int d^4x \sqrt{g}\psi^{\ast}(x)\phi (x)  \label{scpr}
\ee
where $\psi^{\ast}$ represents the complex conjugate. 
Now $D$ is different from the 
usual curved Dirac operator which is denoted by $\stackrel{\circ }{D}$ 
and is defined in 
\cite{lr}, because of the term $K_{\m ab}$ that enters $\o_{\m ab}$ and which 
is required by the supersymmetry. However, $D$ may also have a discrete
spectrum of eigenvalues and eigenvectors. This is not a trivial 
problem and one can think of its possible resolution by considering that 
the operator $\O = i\g^a e_{a}^{\m} \o_{\m bc}(e,\psi ) \g^b \g^c $ 
is a perturbation of $D$ from $\stackrel{\circ }{D}$ and that it  
controls the spectrum of the former. A detailed analysis of this operator 
is not the purpose of this paper and the reader is referred to \cite{vb} for
a deeper discussion of this matter. In what follows it is assumed that 
$D$ admits a discrete spectrum of eigenvalues and 
eigenspinors and thus we can write
\be
D\chi^n = \l^n \chi^n  \label{eigval}
\ee
Then $\l^n$'s define a discrete family on the space of all gravitons and 
gravitinos denoted by $\FF$. As in the case of gravity, $\l^n$'s may 
presumably coordinate the space of orbits of the gauge group in the space 
$\FF$ \cite{vb}. However, even if the above conditions hold, $\l^n$'s might 
fail to be invariant under the gauge group. In fact, the gauge invariance of 
$\l^n$'s impose additional constraints on $\FF$ as well as on the eigenspinors 
of $D$.

To see this, consider the variation of any $\l^n$ under diff's. This variation
is generated by an arbitrary small vector field $\x = \x^{\m}\6_{\m}$. 
The vector field generates infinitesimal transformations 
by the mean of the Lie derivative.
Since
$D=D(e,\psi )$ the eigenvalues $\l^n$ depend on the independent variables 
$e^{a}_{\m}$ and $\psi^{\a}_{\m}$. 
Then we can write for the Lie derivative 
\bea
\d \l^n &=& \frac{\d \l^n }{\d e^{a}_{\m}} \x^{\n} \6_{\n}e^{a}_{\m}+
\frac{\d \l^n }{\d \psi^{\a}_{\m}}\x^{\n} \6_{\n}\psi^{\a}_{\m}\\
&=&<\chi^n | \frac{\d }{\d e^{a}_{\m}} D | \chi^n>\x^{\n} \6_{\n}e^{a}_{\m} +
<\chi^n |\frac{\d  }{\d \psi^{\a}_{\m}}D|\chi^n >\x^{\n} \6_{\n}\psi^{\a}_{\m}
\label{vardif}
\eea
A simple algebra shows that, for the variation (\ref{vardif}) to
vanish, the following set of equations should hold
\be
\TT^{n\m }_{a} \6_{\n}e^{a}_{\m} - \G^{n \m }_{a}\6_{\n}\psi_{\m}^{\a} =0 
\label{difcons}
\ee
where $\TT^{n\m }_{a} = T^{n \m }_{a} +K^{n\m }_{a}$  
where $T^{n\m }_{a}$  
is the "energy-momentum tensor" of the spinor $\chi^n$ 
\cite{lr,dn} 
and
$K^{n\m }_{a}= <\chi^n |i\g_{a}K^{\m }_{bc}(\psi ) \s^{bc}|\chi^n >$.
The last term comes from the derivation of spin-connection 
with respect to the spinor field and is given by
\be
\G^{n \m }_{a} = \frac{i}{4}\int \sqrt{e} {\chi^n}^{\ast}\g^a e_{a}^{\n}
[\bar{\psi_{\n}^{\b}}(\g_{b})_{\a \b }e^{\m}_{c} -
\bar{\psi_{\n}^{\b}}(\g_{c})_{\a \b }e^{\m}_{b} +
\bar{\psi_{b}^{\b}}(\g_{\n})_{\a \b }e^{\n}_{c}]\s^{bc} \chi^{n}. 
\label{gam}
\ee
In a similar manner the invariance of the eigenvalues under the $SO(4)$ raise
new constraints. In this case $e^{a}_{\m}$ transforms under rotations as  
a vector in the upper index while $\psi_{\m}^{\a}$ belongs to the spinorial 
representation of $SO(4)$ ( $\d e^{a}_{\m} = \th^{ab} e_{b\m}, 
\d \psi_{\m}^{\a} =  \th^{ab} (\s_{ab})^{\a}_{\b} \psi_{\m}^{\b} )$ ).
Performing the same steps as in the case of the invariance under diff's 
one gets the following constraints
\be
\TT^{n\m }_{a}e_{b\m} +\G^{n\m }\s_{ab}\psi_{\m}=0 \label{conrot}
\ee
In order to investigate the local susy invariance we consider the following
on-shell local susy transformations
\be
\d e^{a}_{\m} = \frac{1}{2}\bar{\e}\g^{a}\psi_{\m}~~~~~
\d \psi_{\m} = \DD_{\m}\e                \label{susy}
\ee
where $\e(x)$ is an infinitesimal spinor field for which the 
$\bar{\e}=\e^{T}C$ is true. Under (\ref{susy}) the spin connection
transforms as:
\be
\d \o_{\m}^{ab} = A_{\m}^{ab} - \frac{1}{2}e_{\m}^{b}A_{c}^{ac}
+\frac{1}{2}e_{\m}^{a}A_{c}^{bc} \label{trsc}
\ee
where
\be
A_{a}^{\m \n} = \bar{\e}\g_{5}\g_{a}\DD_{\l}\psi_{\r}\e^{\n \m \l \r} 
\label{aterm}
\ee
The vanishing of the variation of the eigenvalues of the Dirac operator
lead to further constraints
\be
\TT^{n\m }_{a}\bar{\e}\g^{a}\psi_{\m} + \G^{n\m }\DD_{\m}\e =0. \label{consusy}
\ee

The set of equations (\ref{difcons}),(\ref{conrot}),(\ref{consusy}) define 
the 
necessary conditions for $\l^n $'s be invariant under the gauge group. These
conditions represent a new type of constraints on the space $\FF$ of all
possible supermultiplets. Furthermore,
as one can see by simply inspecting the relations (\ref{difcons}),
(\ref{conrot}),
(\ref{consusy}) the equations are not independent. Therefore the geometry of
the constrained surface is quite complicated and this makes the quantization
problem highly non-trivial. To deal with this problem one has to work with
BV-BRST or BFV-BRST quantization method developed to handle such situations 
\cite{vb}.

The above constraints are not the only ones arising in this theory. If the 
equation (\ref{eigval}) is subjected to the infinitesimal transformations of 
the gauge group and the variation of $\d \l^n $ vanishes as required 
previously we get 
\be
\d D\chi^n = (\l^n - D)\d \chi^n \label{varr}
\ee
In the case when the above variations are induced by diff's relation 
(\ref{varr}) reads
\be
\{ [b^{\m}(\x ) - c(\l \x )^{\m} ]\6_{\m} + f(\x ) \} \chi^n = 0 \label{dirdif}
\ee
where the following notations are used
\bea
b^{\m}(\x )  &=& i \g^{a} b_{a}^{\m}(\x )~~,~~  
b_{a}^{\m}(\x ) = \x^{\n}\6_{\n}e_{a}^{\m} -e_{a}^{\n}\6_{\n}\x^{\m}
-2e_{a}^{\n}\x^{\m}\o_{\n bc}\s^{bc}\\  
c(\l ,\x )^{\m} &=& (\l^n - D)\x^{\m}~~,~~   
f(\x ) = i\g^{a}\x^{\n}\6_{\n}(e_{a}^{\m}\o_{\m bc})\s^{bc}. 
\eea
A similar relation occurs when in (\ref{varr}) the rotations are considered.
In this case the spin connection transforms as the gauge field for rotations
\be
\d \o_{\m ab} = i [{\bf \th \s}, \o_{\m ab} ] -
i\6_{\m}{\bf \th \s}M_{ab} \label{gaucon}
\ee
where $\th_{ab}=-\th_{ba}$ parameterize an infinitesimal $SO(4)$ rotation and
${\bf \th \s} = \th_{ab}\s_{ab}$. Since $\chi^n$ transforms in the 
unitary spinor representation of $SO(4)$ we can write
\be
\d \chi^n = i {\bf \th\s }\chi^n \label{varchi}
\ee
Using these transformations, the equation (\ref{varr}) becomes
\be
[\th_{a}^{a}D -g(\th ) +h(\th ) ]\chi^n = 0 \label{dirrot}
\ee
where the following notations are used
\bea
g(\th ) &=& [\g^c e_{c}^{\m}([{\bf \th \s},\o_{\m ab}] -
\6_{\m}{\bf \th \s }M_{ab})]\s^{ab} \\
h(\th ) &=&  i (\l^n - D) {\bf \th \s}
\eea
Now if the case of N=1 local supersymmetry is considered, it must be 
noticed that $\chi^n$'s are unaffected by this symmetry and thus the left-hand 
side of equation (\ref{varr}) vanishes. That leads eventually to the equation
\be
[j^{\m}_{a} (\e ) \6_{\m} + k_{a} (\e ) +l_{a} ]\chi^n =0 \label{dirsusy}
\ee
where the notations used above are
\bea
j^{\m }_{a} (\e ) & = & \frac{1}{2}\g_{a}\bar{\e}\psi^{\m} ~,
~~~k_{a}(\e ) = \frac{1}{2}\g_{a}\bar{\e}\psi^{\m}\o_{\m cd}\s^{cd} \\
l_a &=&e_{a}^{\m}[B_{\m cd} - \frac{1}{2}e_{\m d}B_{ec}^{e} +
\frac{1}{2}e_{\m c}B_{ed}^{e}]\s^{cd}.
\eea
The final relations (\ref{dirdif}), (\ref{dirrot}), (\ref{dirsusy}) 
can be interpreted as constraints on the eigenspinors of the Dirac
operator. They depend on the supermultiplets as well as on the eigenvalues
of the Dirac operator and are direct consequences of the invariance of
$\l^n$'s under the gauge group of the problem. These supplementary constraints
complicate the description of the covariant phase space as well as a possible
tentative to formulate the quantum problem in this case \cite{vb}.

In summary, I have discussed the possibility of considering
the eigenvalues of the Dirac operator as observables of Euclidean 
supergravity. The invariance of the eigenvalues under the gauge group of the 
problem imposes severe constraints on the space of the supermultiplets. The 
form of these constraints was completely determined. 
The constraints (\ref{difcons}),(\ref{conrot}),(\ref{consusy}) involve
both the derivatives of the gravitons and the gravitinos as well as
the integral of gravitinos. Since these equations form a system 
non-independent integral-differential equations, at this moment one can only
speculate on their solutions and their form is unknown to the author yet.
However, the relations (\ref{difcons}),(\ref{conrot}),(\ref{consusy})
define a complicate surface on the phase space and thus one has to use BV or 
BFV methods of quantization in order to construct the quantum problem of this
system. 
The eigenspinors get themselves constrained too, from
the requirement that the equations that define the eigenvectors and 
eigenvalues satisfy the gauge group symmetry. The corresponding
equations are (\ref{dirdif}), (\ref{dirrot}), (\ref{dirsusy}) and they
hold whenever the constraints (\ref{difcons}),(\ref{conrot}),(\ref{consusy})
are satisfied, i.e. the gravitons and gravitinos entering these equations 
are the ones obtained as solutions of the constraints.

In the end one must observe that the above considerations are true
only for the case of manifolds with no boundary. Whenever boundary 
hypersurfaces occur, a much larger number of local invariants
can be built and they contribute to the asymptotic expansion of
the integrated heat kernel. This can be used to find appropriate
generalizations of the work by Connes and other authors. At this
moment it is still unclear what kind of boundary conditions (local 
or non-local) form the most appropriate choice in simple or extended 
Euclidean supergravity. The boundary surfaces imply some subtleties
in the case of supersymmetry and one expects that they would affect the 
constraints on the supermultiplets as well as those on the eigenspinors. 

I thank P.A.Blaga for useful discussions and L. Tataru for conversation.

\end{document}